\documentclass[suppldata]{interact}

\usepackage{lmodern}
\usepackage{epstopdf}
\usepackage{caption}
\usepackage{subfigure}

\usepackage{natbib}
\bibpunct[, ]{(}{)}{;}{a}{}{,}
\renewcommand\bibfont{\fontsize{10}{12}\selectfont}

\theoremstyle{plain}

\theoremstyle{definition}

\theoremstyle{remark}

\usepackage{hyperref}
\usepackage[utf8]{inputenc}

\usepackage{booktabs}
\usepackage{longtable}
\usepackage{array}
\usepackage{multirow}
\usepackage{wrapfig}
\usepackage{float}
\usepackage{colortbl}
\usepackage{pdflscape}
\usepackage{tabu}
\usepackage{threeparttable}
\usepackage{threeparttablex}
\usepackage[normalem]{ulem}
\usepackage{makecell}
\usepackage{xcolor}

\begin{document}

\articletype{}

\title{Treatment bootstrapping: A new approach to quantify uncertainty of average treatment effect estimates}

\author{
\name{Jing Li\textsuperscript{a}\thanks{CONTACT Jing Li, Email: jingl8@illinois.edu.}}
\affil{\textsuperscript{a}Department of Political Science, University of Illinois at Urbana-Champaign, IL, USA}
}



\maketitle

\begin{abstract}
This paper proposes a new non-parametric bootstrap method to quantify the uncertainty of average treatment effect estimate for the treated from matching estimators. More specifically, it seeks to quantify the uncertainty associated with the average treatment effect estimate for the treated by bootstrapping the treatment group only and finding the counterpart control group by pair matching on estimated propensity score without replacement. We demonstrate the validity of this approach and compare it with existing bootstrap approaches through Monte Carlo simulation and analysis of a real world data set. The results indicate that the proposed approach constructs confidence intervals and standard errors that have 95 percent or above coverage rate and better precision compared with existing bootstrap approaches, while these measures also depend on percent treated in the sample data and the sample size.
\end{abstract}

\begin{keywords}
Pair-matching; ATT; Non-parametric Bootstrap; Propensity Score
\end{keywords}

\hypertarget{introduction}{%
\section{Introduction}\label{introduction}}

Applications of causal inference methods have increased dramatically in the last few decades. In observational studies, because of the non-random assignment of treatment, the treatment and control groups too often are unbalanced on their observed covariates. One way to remedy this problem is through matching \citep{Rosenbaum83, Rosenbaum85, Abadie11, Diamond13, Sekhon11}. While a lot of scholarly attention has duly been focused on estimating the average treatment effect (ATE) using different matching estimators, the equally important issue of the uncertainty of average treatment effect estimates has been emphasized much less. This is reflected in the fact that there is a fair amount of variation and ambiguity in how applied causal inference research derives and reports uncertainty of ATE estimates. While most studies do report standard error or confidence interval \citep{Aldrich11, Boas11, Dehejia02, Galiani05, Green05, Hong16, Kam08, Kocher11, Mayer11, Mozer20, Simmons05, Urban14}, only a minority of these ones mention which specific method is used \citep{Dehejia02, Galiani05, Green05, Mayer11, Urban14}.

This often leaves readers unsure how uncertainty measures of ATE estimates are derived. And even if the name of a particular method is mentioned, the bootstrap method for example, readers are still unsure of the specific steps taken to form the uncertainty measures. In this paper, we explore uncertainty measures for ATT estimates and more specifically uncertainty measures derived from the non-parametric bootstrap method.

While the non-parametric bootstrap method and its rich array of variants and extensions are common ways to derive standard error or confidence interval for ATE estimates especially for matching estimators, a closer scrutiny of this method in practice is still much needed. In particular, in this paper, we propose a new way of quantifying uncertainty of average treatment effect for the treated (ATT) estimates, compare it with existing methods, and discuss their relative strengths and weaknesses. While the focus in this paper is on ATT estimates, the proposed method can be applied to a broad set of causal effect estimators. 

The paper is organized as follows: section 2 briefly reviews the potential outcome framework for causal inference; section 3 discusses the uncertainty of ATE estimates; section 4 reviews relevant existing bootstrap approaches and provides details on the proposed bootstrap approach for deriving standard error and confidence interval of sample ATT estimates; section 5 presents simulation studies and results; section 6 offers discussions and conclusions.

\hypertarget{causal-inference-and-causal-effect-estimation}{%
\section{Causal Inference and Causal Effect
Estimation}\label{causal-inference-and-causal-effect-estimation}}

The fundamental problem in causal inference is that we cannot observe a unit in its factual and counter-factual states at the same time. Therefore, for two potential outcomes $Y_i(1)$ and \(Y_i(0)\), a unit can only be either treated or not treated but not both at the same time \citep{Holland86}. Obtaining the counter-factual state for either treated or control units necessitates separating out the effect of all factors on the potential outcomes other than that of the treatment. The Neyman-Rubin causal model shifts the problem of causal effect identification and inference to estimating the average treatment effect (ATE) between a pair of comparable treatment and control groups \citep{Rubin74, Rubin78}. That is: 
\begin{equation}
ATE = E(Y_i(1) - Y_i(0)) = E(Y_i \mid T = 1) - E(Y_i \mid T = 0)
\end{equation}

Randomized experiment treatment assignment balances both observed and unobserved covariates between the treatment and control groups, thus making the treatment and control groups comparable, obtaining counter-factual states for both treatment and control groups, and enabling estimation of the average treatment effect as a result. That is:
\begin{equation}
ATE = E(Y_i \mid T = 1) - E(Y_i \mid T = 0) = \frac{1}{n_t}\sum_{i=1}^{n_t} Y_{i} - \frac{1}{n_c}\sum_{i=1}^{n_c} Y_i
\end{equation}

When randomized treatment assignment doesn't exist as in observational settings, statistical adjustment methods such as matching are needed to account for the non-random assignment of treatment, identify the counter-factual states, and enable estimation of the average treatment effect\citep{Diamond13, Hansen04}. More specifically, we may estimate the average treatment effect for the treated (ATT) and the average treatment effect for the control (ATC):
\begin{equation}
ATT = E((Y_i(1) - Y_i(0))|T = 1) = E(Y_i(1)|T=1) - E(Y_i(0)|T=1)
\end{equation}
\begin{equation}
ATC = E((Y_i(1) - Y_i(0))|T = 0) = E(Y_i(1)|T=0) - E(Y_i(0)|T=0)
\end{equation}

When treatment assignment is randomized, potential outcomes have statistical independence with treatment assignment, and \(ATE = ATT = ATC\), otherwise \(ATE\) does not equal \(ATT\) or \(ATC\) in general. In this paper, we focus on \(ATT\).

\hypertarget{uncertainty-of-causal-effect-estimation}{%
\section{Uncertainty of Causal Effect
Estimate}\label{uncertainty-of-causal-effect-estimation}}

While the uncertainty of treatment effect estimates has been acknowledged by statisticians since the early days in the development of causal inference framework\citep{Fisher35, Neyman23}, in applied studies, researchers seem to have a tendency to gloss over this important component of the causal inference process. In applied settings, while it is common that standard error or confidence interval of causal effect estimate is provided, often times no details are presented on exactly how these quantities are derived \citep{Abadie03,Aldrich11, Boas11, Christakis03, Dehejia02, Diamond13, Galiani05, Green05, Hansen04, Hong16, Kam08, Kocher11, Mayer11, Mozer20, Simmons05, Urban14}. 

In a number of studies, names of the procedure used, for example 'bootstrap', are mentioned \citep{Dehejia02, Galiani05, Green05, Mayer11, Urban14}, but again no details are given. This often leave readers wonder precisely how the standard errors or confidence intervals are generated. In a few other works, uncertainty of causal effect estimates are not addressed at all \citep{Abadie10, Henderson11}. Notable exceptions are \citet*{Rosenbaum02} which constructs confidence interval of treatment effect estimates by inverting hypothesis tests and collecting all values not rejected at a certain \(\alpha\) level into a confidence set, and \citet*{Abadie02, Imbens04} which offer an analytic solution for deriving the variance of ATE estimates.

Therefore, developing the best practice for quantifying the uncertainty of ATE estimates is still much needed and it is just as important as estimating ATE in the first place. And while in certain cases researchers have derived mathematical expressions for variance estimators for certain specific ATE estimators \citep{Imbens2015, Rojas2009}, these analytical solutions do not serve more general use cases and often are not intuitive to understand. 

Then, how can we quantify the uncertainty associated with ATE estimates? In discussing the different approaches used for deriving confidence interval for treatment effect estimates from propensity score matching: normal-theory approaches, resampling approaches, and randomization-based approaches, \citet*{Hill06} shows that bootstrap-based methods have superior performance compared with other methods in certain situations. In the following, we also propose a new non-parametric bootstrap approach for measuring uncertainty associated with ATT estimates.

\hypertarget{non-parametric-bootstrapping-of-average-treatment-effect-for-the-treated-att}{%
\section{Non-parametric Bootstrapping of Average Treatment Effect for
the Treated
(ATT)}\label{non-parametric-bootstrapping-of-average-treatment-effect-for-the-treated-att}}

\subsection{Existing bootstrap approaches for quantifying uncertainty of ATE estimates}

The bootstrap method as first introduced by Efron \citep{Efron1979} is a non-parametric approach to quantify the uncertainty of parameter estimates when there is not much information on the population distribution of parameter of interest. The key principle is to re-sample from an original data sample with replacement, which produces replicated resample data with desired random variation embedded within, thus enabling researchers to form uncertainty measures such as standard error and confidence interval. The rationale behind the bootstrap method is that the bootstrap distribution of \(\tilde\tau - \hat\tau\) closely mimics the sampling distribution of \(\hat\tau - \tau\), known as the bootstrap Central Limit Theorem \citep{Singh81}.

However, bootstrap is not a one-fits-all method and it can fail for certain types of data and for certain types of statistics\footnote{The formal conditions where the bootstrap generally works are: for any distribution \(A\) within a neighborhood of the true distribution F, the bootstrap distribution \(G_{A, n}\) converges weakly to a limiting distribution \(G_{A, \infty}\); this convergence is uniform on the neighborhood; and the mapping from \(A\) to \(G_{A, \infty}\) must be continuous \citep{Bickel81, Davison97}.}. Cases where the simple bootstrap can fail include when the statistic of interest is the sample maximum, unstable or unsmooth statistic, for example, the sample median \citep{Bickel81, Davison97}, or when the parameter lies on the boundary of the parameter space \citep{Andrews2000}; or when the sample data in question has dependence structure, for example, auto-correlated time series data, incomplete data, or dirty data \citep{Davison97}. Therefore, when leveraging the bootstrap method to quantify uncertainty associated with ATE estimates, cautions ought to be observed.

For matching estimators more specifically, \citet*{Abadie08} show that the standard bootstrap in general is not valid for simple nearest-neighbor matching with replacement and a fixed number of neighbors \footnote{They do point out that for asymptotically linear estimators such as propensity score pair-matching without replacement, the bootstrap method does provide valid inference.}, while they provide analytical asymptotic variance estimators for matching with replacement and a fixed number of matches \citep{Abadie06} and propensity score matching with replacement \citep{Abadie16}. \citet*{Abadie22} shows that a matched bootstrap, which resamples matched sets of one treatment unit and k control units without replacement, does provide valid standard error for ATE estimates.

In addition, \citet*{Bodory20} shows that a wild bootstrap procedure, which is based on a martingale representation of matching estimators proposed by \citet*{Abadie12}, performs well compared with standard bootstrap and (conservative) asymptotic variance approximations. \citet*{Otsu17} proposes a weighted bootstrap approach by directly taking the number of times a unit is used for matching as an attribute of the unit when bootstrapping \footnote{This inference method is for matching with replacement.}. \citet*{Austin14} show the validity of a paired bootstrap procedure which bootstraps pairs of matched treatment and control units. Our proposed approach follows this line of work, and more specifically bootstraps the sample treatment group first and then construct the counterpart control group by pair matching on estimated propensity score without replacement, and does not contradict with the results of \citet*{Abadie08}.

\subsection{A comparison of different bootstrap approaches}

Below we make a comparison between the proposed approach and three most commonly used bootstrap procedures for deriving uncertainty measures of average treatment effect estimates. 

The first kind of bootstrap is to bootstrap the sample treatment group and control groups separately \citep{Hall88, Tu02, Abadie08}; the second one is a paired bootstrap which bootstraps pairs of treatment and control units after performing pair matching on estimated propensity score without replacement on the sample data \citep{Austin14}; the third and most common one is to bootstrap the whole sample of both treatment and control units together \citep{Austin17, Austin22, Cerulli14, Bodory20} \footnote{We specifically refer to the standard bootstrap in \citep{Bodory20}. Other than these three bootstrap approaches listed, there are possibly a lot more non-parametric bootstrap methods that have been used to derive uncertainty measures of average treatment effect estimates for matching estimators. However, the focus of this paper will be on these three approaches as they are easy to understand and comparison between the proposed approach and these three basic approaches can shed light on the statistical properties of other variants of bootstrap uncertainty measures. In addition, as most applied research using the bootstrap method to quantify uncertainty of ATE estimates does not report exactly how the bootstrap is done, chances are there are many more applied works that use the three basic bootstrap approaches listed above in addition to the ones cited here.}. In this paper, we only consider matching without replacement and compare the three bootstrap approaches listed above with the proposed approach detailed below.


\citet*{Embens21} differentiates between uncertainty from sampling variation and uncertainty from the stochastic nature of the treatment assignment. To adjudicate among the four different bootstrap approaches, it is also helpful to examine what kind of uncertainty they are trying to identify and estimate. The paired bootstrap approach by \citet*{Austin14} resamples pairs of matched treatment and control group units. It can be argued that a pair of treatment and control group units represents an individual treatment effect estimate, that is \citet*{Austin14} is essentially bootstrapping individual treatment effects. Therefore, the main uncertainty accounted for by the paired bootstrap approach is how individual treatment effect could vary across the sample treatment units. The main drawback of this approach is that for each bootstrap iteration, it does not redo the matching process. The matched set bootstrap approach in \citet*{Abadie22} can be seen as an extended version of the paired bootstrap where the number of control units matched to a treatment unit will not necessarily be 1.

When bootstrapping sample treatment and control groups separately, we are also accounting for the uncertainty associated with the treatment group and that of the control group separately. The potential drawback of this approach is that in an observational setting, there could be significant overlaps between uncertainty associated with the treatment group and that of the control group. And accounting for the two sources of uncertainty separately could potentially overestimate the uncertainty associated with sample ATT estimates. 

When bootstrapping both the sample treatment group and the sample control group together, we are accounting for the uncertainty associated with the sample data as a whole. However, this uncertainty may not necessarily correspond to the uncertainty associated with the ATT estimate. In one bootstrap step, the number of treatment group units, the number of control group units, the composition of the treatment group, and the composition of the control group all change at the same time while the sample ATT estimate does not necessarily change in a similar fashion \footnote{Again \citet*{Embens21} characterizes two types of uncertainty and bootstrapping the whole sample data cannot account for both at the same time.}. In real world settings, the simplest way to think about how treatment effect can vary is to form many randomized experiments\footnote{More specifically, in an real world experimental setting, we would expect the people who receive treatment will change with each experiment, and the corresponding control group also changes such that it matches the treatment group (on both observed and unobserved covariates) as the pair of treatment and control groups forms a randomized experiment.}. And in observational settings, to mimic the construction of many experiments, the most straightforward way is to create many matched pairs of treatment and control groups. 

The question is how best to construct many matched pairs of treatment and control groups to mimic randomized experiments as close as possible with observational data. Here, we take a very specific perspective on the uncertainty associated with ATT estimates. For matching estimators of ATT more specifically, our starting point is the sample treatment group. Therefore, to accurately identify and estimate the uncertainty of ATT estimates, we start with accounting for the uncertainty associated with the sample treatment group and then account for the uncertainty associated with the subsequent matching process and more specifically the uncertainty associated with the matched control group. To account for the uncertainty associated with the sample treatment group, we just perform a simple bootstrap step. And to account for the uncertainty associated with the matched control group, we just pair-match control group units with bootstrap resample treatment group units on estimated propensity score\footnote{This does not necessarily require the sample ATT matching estimator to be the propensity score matching estimator.}. While we are not explicitly accounting for the sampling variation and treatment assignment variation as outlined by \citet*{Embens21}, we are accounting for the same two sources of uncertainty involved in the matching process in a different and valid way. 

We focus on ATT because when we try to identify and estimate the average treatment effect, the expectation is that the treatment in question will \textbf{have} an effect on the outcome for those that are treated and \textbf{not have} an effect for those not treated. Therefore, we are most interested in how the average treatment effect can vary for those that are treated. In this sense, no wonder that ATT is the most commonly used causal estimand. 

The proposed approach also remedies a problem commonly encountered in bootstrap inference for causal effect estimate \citep{Abadie08, Otsu17}, that is units in the sample data appear only once while a bootstrap resample treatment group will almost surely contain certain sample data units more than once. Therefore, if we are to find the exact matched counterpart control group for a bootstrap treatment group resample such that these two groups mimic a randomized experiment as close as possible, we will not be able to find any because the sample control group doesn't have any units that appear more than once. Our proposed approach to match bootstrap resample treatment group with sample control group on estimated propensity score remedies this issue. A pair of bootstrapped treatment group resample and its matched counterpart control group serves as a close approximation to a randomized experiment if certain assumptions as outlined in section ~\ref{validity} are met\footnote{In social demographic settings, individuals subject to a treatment usually is a small group while the potential control pool is quite large, therefore, in general, there is a large number of control units available to match with the treatment group units as demonstrated below in a simulation using the CPS data set.}.  

\subsection{A new non-parametric bootstrap uncertainty estimator for matching estimators}

The detailed procedure of how we derive the bootstrapped ATT estimates (for matching estimators) is as follows:

\textbf{Non-parametric Bootstrapped ATT Estimate:}

\textbf{Step 1}. Calculate a propensity score for each of the units in the sample data set using the logit model;

\textbf{Step 2}. Match a set of units from the sample control group to the sample treatment group based on estimated propensity score from step 1 without replacement;

\textbf{Step 3}. Compute the ATT estimate by calculating the mean difference of the outcome $Y$ between the treatment group and the matched control group;

\textbf{Step 4}. Bootstrap the treatment group (sample with replacement\footnote{Same size as the original sample treatment group.}) 500 times to obtain 500 bootstrap resamples, repeat steps 2 and 3 to find matched control group counterpart for each bootstrap treatment group resample\footnote{This ensures that the matched counterpart control group for each bootstrap treatment group resample is dynamic. It is superior to the approach of bootstrapping separately the sample treatment group and the sample control group, because here the bootstrap procedure deals with one source of uncertainty that is easy to quantify while bootstrapping both treatment and control groups separately tries to simultaneously account for two sources of uncertainties that are difficult to quantify and may lead to misleading estimates.}, and calculate an ATT estimate for each pair of bootstrap treatment group resample and its matched control group counterpart.

\subsection{Validity of the proposed bootstrap method}\label{validity}

Consider a basic setup where we have $X, Y, Z$ denoting different random variables. $Z$ is a binary treatment indicator, for example, participation in a job training program, $Z = 1$ for the treatment group and $Z = 0$ for the control group. $Y$ is the outcome variable, $X$ refers to the $p \times 1$ set of covariates for both the treatment and control groups (for example, demographic variables such as race, education, income, gender, age, etc.). The question of interest is to estimate the effect of the treatment on the outcome for units in the treatment group.  

Let us define $\tau$ as the population statistic of interest, namely average treatment effect for the treated, the sample ATT estimate as $\hat \tau$, and the bootstrap resample ATT estimate as $\tilde\tau$. For the bootstrap method, the assumption is that the distribution of $\tilde \tau - \hat \tau$ can approximate the distribution of $\hat \tau - \tau$ well. Suppose the sample size of the matched sample is $n$. If we assume $\sqrt n (\hat\tau - \tau) \to_d N (0, \sigma^2)$ for some $\sigma > 0$, then $\sqrt n (\tilde\tau - \hat\tau) \to_d N (0, \sigma^2)$. 

As is common in the literature, we adopt a few assumptions that will enable us to establish the validity of the proposed bootstrap approach. 

\emph{Assumption 1 (random sampling)}: sample data $S = \{ Y_i, X_i, Z_i \}^N_{i = 1}$ consists of $N_1$ treatment and $N_0$ control units, which are random draws from the population distribution $(Y, X)$, $N = N_1 + N_0$. And let $S^\ast \subseteq S$ be the matched sample obtained via pair-matching without replacement on estimated propensity score $\gamma_i = f(Z_i, X_i)$.

\emph{Assumption 2 (common support condition)}: let $\Gamma_1 = supp(\gamma | z = 1)$, and $\Gamma_0 = supp(\gamma | z = 0)$, then $\Gamma_1 \subseteq \Gamma_0$. This ensures that enough good matches from the control group can be found for the treatment group units.

\emph{Assumption 3 (matching discrepancies)}: with $N_0 >> N_1$, the assumption is that regardless of which specific matching algorithm being used, the procedure will be able to ensure that $X_1$ and $X_0$ of the matched sample have close enough distributions.

Matching as a type of quasi-experimental design creates a matched sample $S^\ast$, which can mimic a randomized experiment and provide a treatment effect estimate. This estimate asymptotically is expected to recover the true treatment effect in the target matching population $P^\ast$, in which the treatment and control groups have the same distribution of $X$. That is ATT estimate $\hat\tau$  from the matched sample asymptotically can recover the true treatment effect $\tau$ in the population. 

\emph{Assumption 4 (conditional independence)}: the assumption is that conditional on observed covariates $X$, the potential outcome $Y(1)$ and $Y(0)$ is independent of the treatment assignment, $(Y(1), Y(0)) \perp Z$ \citep{Rosenbaum83}. However, in order for this to hold, we need to assume that no unobserved confounding exists that can systemically bias the relationship between $Y$ and $Z$.

Let the variance of the bootstrap ATT estimate $\tilde \tau $ be $\lambda_n(\tilde\tau)$, and $n = 500$, then the bootstrap procedure is said to be valid if $\lambda_n(\tilde\tau) - \lambda_n(\hat \tau) \to_p 0$. With the random sampling assumption above and a large enough sample size $n = 500$, it is typically the case that $\lambda_n(\hat \tau) \to \lambda(\tau)$. Then to prove that $\lambda_n(\tilde\tau) - \lambda_n(\hat \tau) \to_p 0$, we only need $\lambda_n(\tilde\tau) \to_p \lambda (\tau)$. And with $\lambda_n(\tilde\tau) - \lambda_n(\hat \tau) = (\lambda_n(\tilde\tau) - \lambda(\tau) ) - (\lambda_n(\hat \tau) - \lambda(\tau)) $, we can prove that $\lambda_n(\tilde\tau) - \lambda_n(\hat \tau) \to_p 0$ \footnote{Let $G_n(t) = \lambda_n(\hat \tau) = P(\sqrt n (\hat \tau - \tau) \le t), ~ \tilde G_n(t) = \lambda_n(\tilde \tau) = P(\sqrt n (\tilde \tau - \hat \tau) \le t)$, using the Berry-Essen theorem, it can be shown that $\underset {t} {\text{supp}} |\tilde G_n(t) - G_n(t)| = O_p(1/\sqrt n)$, therefore, the bootstrap approach is statistically valid.}. 

Below we demonstrate the validity of the proposed bootstrap approach through a Monte Carlo simulation and a simulation using the Current Population Survey (CPS) data set, and compare it with the main existing alternatives. 

\hypertarget{analysis}{%
\section{Simulation Studies}\label{analysis}}

\hypertarget{results}{%
\subsection{Monte Carlo Simulation}\label{Monte Carlo Simulation}}

Here we perform Monte Carlo simulation to examine the behavior of our proposed bootstrap ATT uncertainty estimator and make a comparison with the three existing alternative approaches. 

In accordance with the random sampling assumption, our sample data are random samples from a super population of size 1000000. The sample size of a random sample from the super population can differ, and we have three different values: 500, 1000, 2000. In addition, the percentage of treatment units in the super population varies across six different scenarios: 5 percent, 10 percent, 15 percent, 20 percent, 25 percent, 30 percent. For each of these six scenarios, random samples of size 500, 1000 and 2000 are generated. 

The data generating process of the super population mimics the ones used in the literature for a continuous outcome \citep{Austin14, Austin22}. There are 10 baseline covariates: $X_1, ..., X_{10}$. These covariates all follow the standard normal distribution. Seven of them $X_1, ..., X_7$ affects the treatment assignment and seven of them $X_4, ..., X_{10}$ affects the outcome. And the covariates' effect on the treatment assignment or the outcome could be weak, medium, strong, or very strong. The probability of treatment is formed from a logit model: $\text{logit}(p_i) = \alpha_{0, \text{treat}} + \alpha_w x_1 + \alpha_m x_2 + \alpha_s x_3 + \alpha_w x_4 + \alpha_m x_5 + \alpha_s x_6 + \alpha_{vs} x_7$. And the treatment status is generated from a Bernoulli distribution with unit-specific parameter $p_i: Z_i \sim \text{Bernulli}(p_i)$. The continuous outcome was generated from the following formula: $Y_i = Z_i + \alpha_w x_4 + \alpha_m x_5 + \alpha_s x_6 + \alpha_{vs} x_7 + \alpha_w x_8 + \alpha_m x_9 + \alpha_s x_10 + \epsilon_i$, where $\epsilon_i \sim N(0, \sigma = 3)$. The true treatment effect is 1, that is the treatment increases the average outcome by 1. 

\begin{figure}[!tbp]
  \centering
  \includegraphics[width=\textwidth]{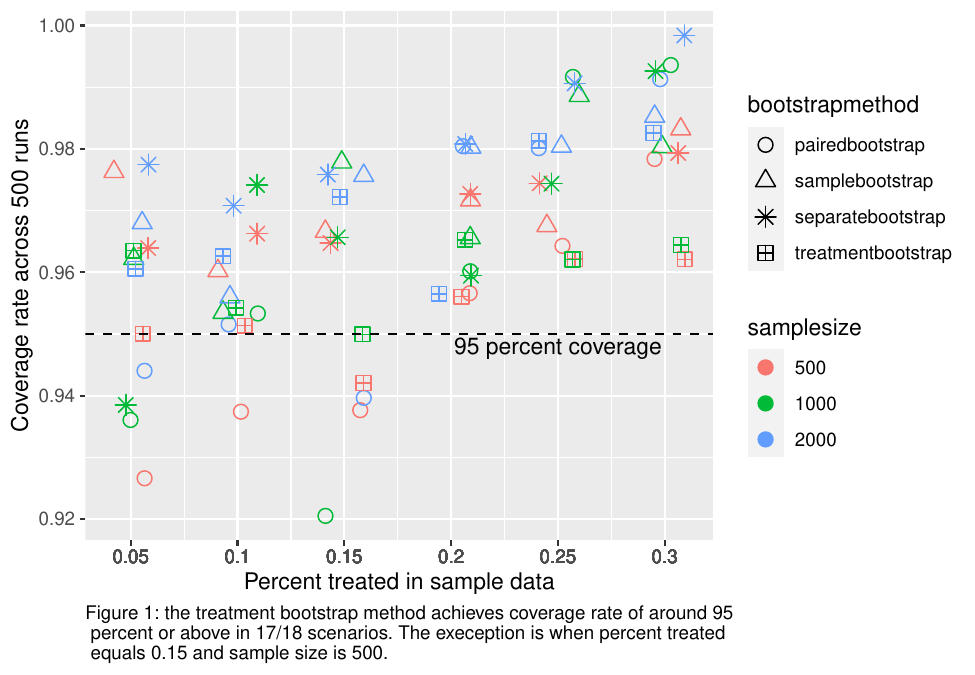}
\end{figure}

For each of the sample data, we perform propensity score matching using optimal matching (with a covariate distance tolerance level of 0.02). The same matching procedure is applied for the 18 different data scenarios (6 different percent treated in sample data and 3 different sample size) and 4 different bootstrap methods for deriving ATT uncertainty measures. And the main statistical properties of ATT uncertainty measures we look at are coverage rate and average standard error across 500 iterations for each scenario. The coverage rate refers to the number of times the 95 percent confidence interval (\(\tau - 1.96 se_B, \tau + 1.96 se_B\)) traps the true treatment effect value 1, and the average standard error is the arithmetic mean of 500 estimated standard errors (\(se_B(\tilde\tau) = \sqrt{\frac{1}{n}\sum_{i=1}^n(\tilde\tau_i - \hat\tau)^2}\)) calculated from 500 runs for each data scenario and each bootstrap method. We also examine the covariate balance of the matched samples in terms of maximum absolute standardized mean difference, thus making sure that the matching procedure achieves the goal of balancing treatment and control groups in terms of observed covariates. 


In Figure 1, we can observe that in 17 out of 18 data scenarios, the proposed approach treatment group bootstrap approach achieves coverage rate of 95 percent or above. The exception is when percent treated in the sample is 0.15 and sample size is 500. Even then the coverage rate is about 0.94. Overall, the proposed bootstrap approach achieves satisfactory coverage rate.

\begin{figure}[!tbp]
  \centering
  \includegraphics[width=\textwidth]{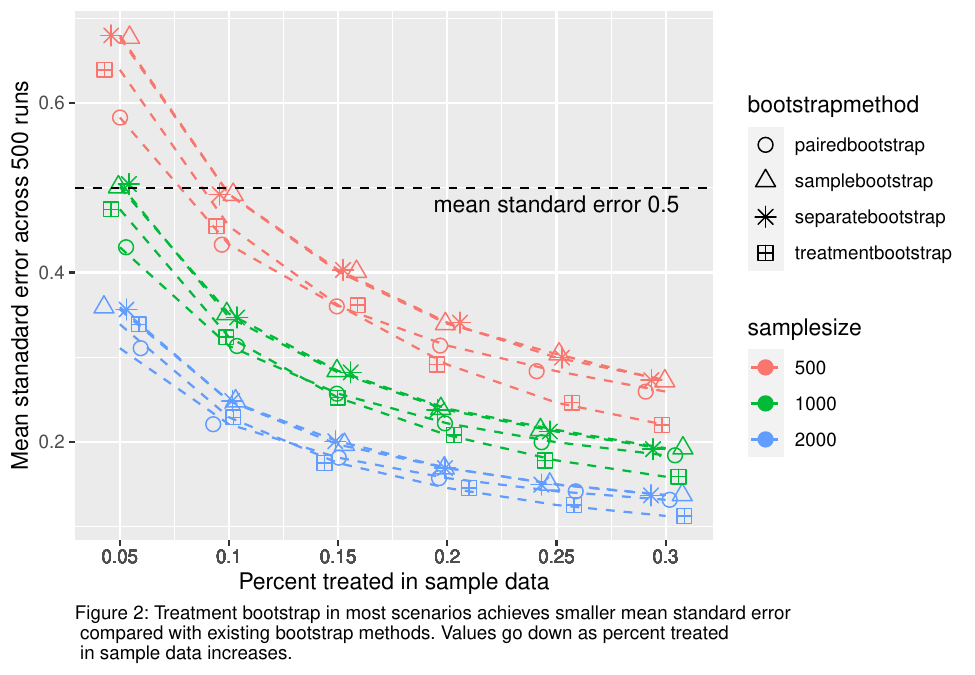}
\end{figure}

\begin{figure}[!tbp]
  \centering
  \includegraphics[width=\textwidth]{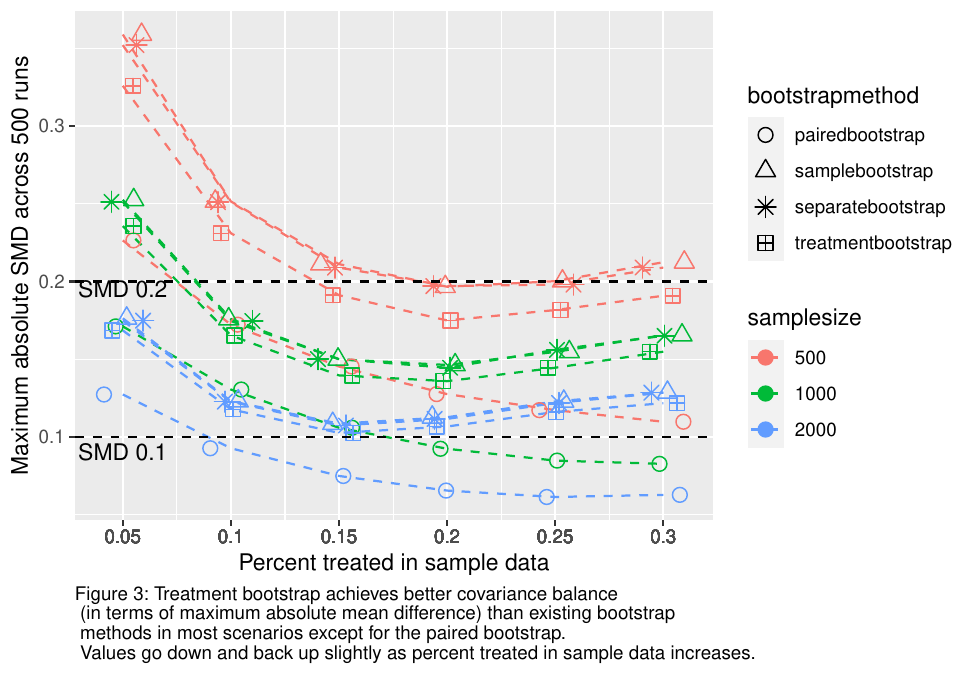}
\end{figure}

In Figure 2, we can observe that in 17 out of 18 scenarios, the proposed bootstrap approach achieves average standard error lower than 0.5, the value that makes zero 2 standard errors away from 1, the true value of ATT. And as the percent treated in the sample increases, the average standard error decreases for all bootstrap methods while the proposed treatment group bootstrap method achieves lower mean standard error compared with the three existing bootstrap approaches. A large sample size also makes average estimated standard error smaller, and in particular, the proposed bootstrap approach achieves the lowest average standard error of around 0.1 when percent treated in the sample is 0.3 and the sample size is 2000.

In Figure 3, we can observe that the maximum absolute standardized mean difference (MASMD) for the proposed bootstrap approach is comparable or smaller than the sample and separate bootstrap approaches but larger than the paired bootstrap approach. As aforementioned, the paired bootstrap approach does not redo matching in each bootstrap iteration, therefore, it is no surprise that paired bootstrap has the smallest maximum absolute standardized mean difference. Among the methods that do redo matching in each bootstrap iteration, the proposed treatment bootstrap approach has comparable or better covariate balance. In addition, MASMD for the proposed approach is mostly between 0.1 and 0.2, two thresholds that are widely used to determine post-matching covariate balance. The exceptions are when percent treated is 0.05 and the sample size is either 500 or 1000 and when percent treated is 0.1 and sample size is 500.

For the three bootstrap approaches other than paired bootstrap, MASMD shows a slight upward trend when percent treated is between 0.2 and 0.3. The reason is that as percent treated increases, the size of the overlap in estimated propensity score between the bootstrap treatment group and the sample control group increases while the average number of potential matches for treated units decreases at the same time. Apparently starting from percent treated 0.2, the decrease in average number of potential matches outweigh the increase in overlap in estimated propensity score \footnote{When sample size is 2000, this increase in MASMD starts earlier than when sample size is 500 or 1000.}. Corresponding plots for the size of the overlap region and the mean number of potential matches for the treated units are shown in the Appendices A and B.

This provides numerical evidence that the proposed bootstrap approach forms comparable or more precise uncertainty measures for ATT estimates. The corresponding coverage rates are quite satisfactory, in 17 out of 18 cases reaching 95 percent or above. Meanwhile, the corresponding bootstrap treatment group resample and matched control group also has sound covariate balance. Both per treated in the sample and sample size can have sizable influence on these three quantities.

\hypertarget{results}{%
\subsection{Simulation with the CPS dataset }\label{Simulation using the CPS dataset}}

\begin{figure}
\centering
\includegraphics[keepaspectratio=true, height = 9 cm]{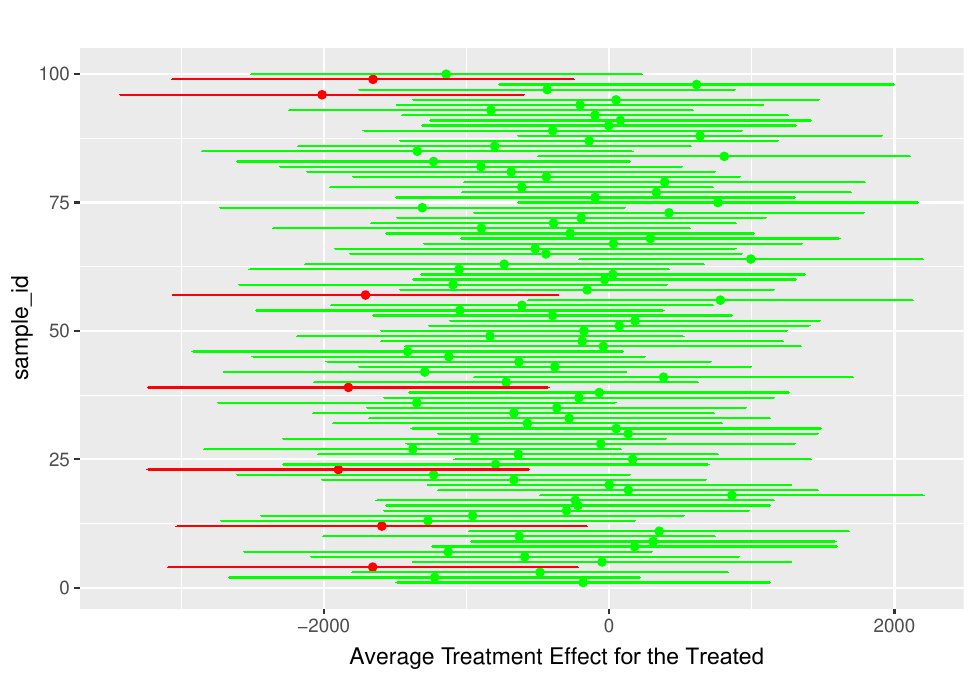}
\caption*{Figure 4: the coverage of the proposed bootstrap approach for ATT estimate using the CPS data set. 93 out 100 confidence intervals trap the true treatment effect 0, 3 of the other 7 are pretty close to have the true treatment effect 0 within 2 standard deviations.}
\end{figure}

\begin{figure}
\centering
\includegraphics[keepaspectratio=true, height = 9 cm]{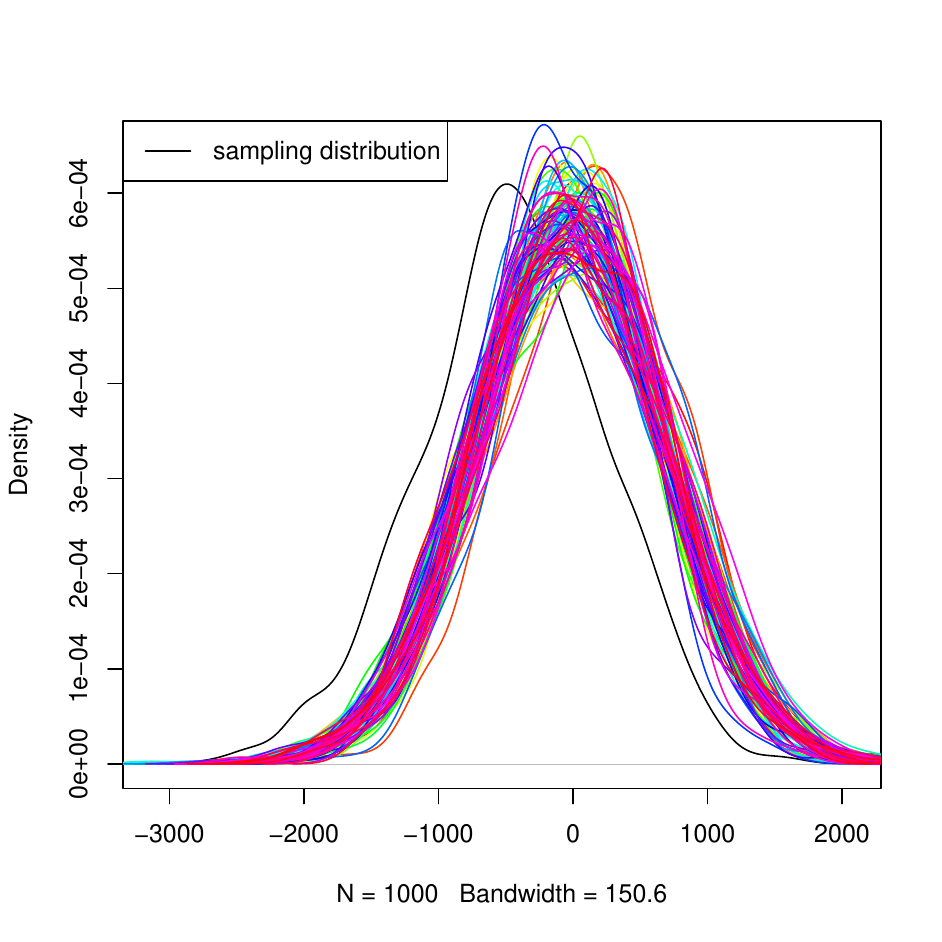}
\caption*{Figure 5: the sample distribution of the ATT estimates and 100 bootstrap distribution of ATT estimates. The bootstrap ATT distributions center around 0, the true treatment effect. The sample ATT distribution centers slightly to the left of 0.}
\end{figure}

Different from the Monte Carlo simulation above, here we examine the behavior of the proposed bootstrap approach as well as the three existing ones using a real-world data set, the Current Population Study (CPS) data set. First, we randomly assign 175 individuals \footnote{This treatment group size mimics the Dehejia-Wahha Sample of the Lalonde NSW experiment data \citep{Lalonde86}.} to be the treatment group while the remaining units act as the control group. Repeating this exercise 1000 times gives us 1000 simulated data sets. Applying propensity score matching to each of these 1000 data sets, and we can obtain 1000 pre-bootstrapped ATT estimates. This is the sampling distribution of ATT. As all units in CPS data set are not treated, the average treatment effect estimate should be zero in expectation\footnote{That is the true treatment effect is zero.}. The sample mean of the non-bootstrapped ATT is -450.9568 (real earnings in dollars in 1978), which is pretty close to the true treatment effect zero while the standard deviation is 685.1427.

Results displayed in Figure 4 show that the proposed non-parametric bootstrap procedure can quantify the uncertainty of ATT estimates reasonably well. Across 100 runs of the proposed bootstrap procedure for 100 samples (randomly drawn from the 1000 sample data sets generated above), 93 out of the 100 bootstrap confidence intervals trap the true treatment effect of 0, while in 3 of the other 7 cases, the true treatment effect 0 is pretty close to be within two standard errors of the ATT estimate. The green error bars in Figure 4 are those 93 which have 0 within 2 standard errors away from the sample ATT estimate while the red error bars are those 7 which do not have 0 within 2 standard errors away from the sample ATT estimate.

\begin{figure}
\centering
\includegraphics[keepaspectratio=true, height = 9 cm]{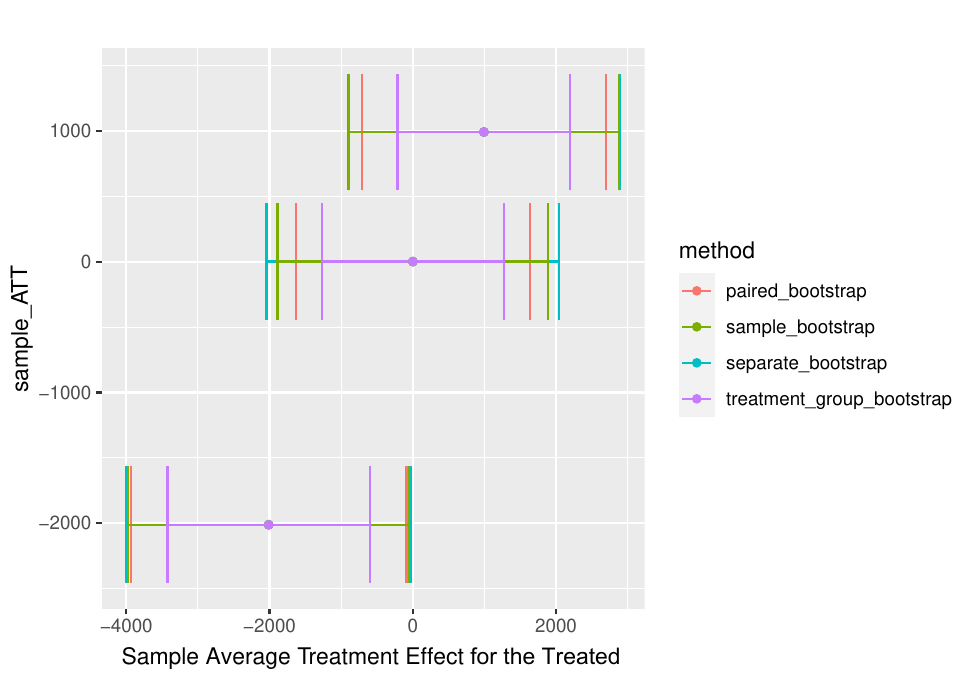}
\caption*{Figure 6: comparison of different bootstrap confidence intervals for three different sample ATT estimate scenarios: sample ATT smallest, largest, closet to the true treatment effect 0. The proposed approach shows narrower confidence intervals in all cases.}
\end{figure}

In Figure 5, we plot the sample distribution of ATT, the single black curve on the left, that is the distribution of ATT estimates of the 1000 non-bootstrapped samples, along with 100 bootstrap distributions of ATT, the 100 colored curves on the right, the 100 data samples for which we calculated bootstrapped standard error and 95 percent confidence interval of the sample ATT estimate. It can be observed that the bootstrap distributions are in same shape as the sample distribution of ATT, the only difference is that the sample distribution is centered slightly to the left.

In Figure 6, we show the approximate 95 percent confidence intervals for the proposed bootstrap method as well as the three existing ones for three difference cases of sample ATT: sample ATT has the smallest negative value; is closest to zero (the true treatment effect); has the largest positive value. It can be observed that in all three cases, the proposed approach achieves the narrowest confidence interval. Therefore, the uncertainty measure derived from the proposed bootstrap approach appears to be more precise than the existing ones. 

\hypertarget{discussions-and-conclusion}{%
\section{Discussions and Conclusion}\label{discussions-and-conclusion}}

Compared with existing non-parametric bootstrap approaches, the proposed approach of bootstrapping the treatment group only and finding the counterpart control groups by pair matching on estimated propensity score without replacement has shown to be able to provide more precise uncertainty measures for ATT estimates in most scenarios in the Monte Carlo simulation and the simulation using CPS data set. It also provides satisfactory coverage rates of 95 percent or above in vast majority of data scenarios. Moreover, covariate balance between bootstrap treatment group resample and matched control group is sound. Nonetheless, it is necessary to be aware that both per treated in sample data and sample size can have an influence on the properties of uncertainty measures for ATT estimates.

In essence, our proposed bootstrap approach tries to account for the two sources of uncertainty involved in estimating ATT in a specific but intuitive way. First, we deal with the uncertainty associated with the treatment group by simply bootstrapping the treatment group, and then we deal with the uncertainty associated with the matched control group by pair matching the bootstrap treatment group resamaple with the sample control group on estimated propensity score without replacement. A pair of bootstrap treatment group resample and its matched control group counterpart effectively serves as an approximation to a randomized experiment.

The proposed approach of dealing with uncertainty involving multiple groups has the potential to be applied to a broad set of causal effect estimators. For ATC estimates, we would first bootstrap the sample control group, match the bootstrap resample control group with the sample treatment group, and then calculate the ATC estimate for each pair of bootstrap resample control group and the matched treatment group. For ATT estimates from weighting estimators, we would first bootstrap the treatment group and apply the weighting procedure with the sample control group, and calculate the ATT estimate for each pair of bootstrap treatment group resample and the weighted sample control group. 

In the simulations above, we match individual units in each sample data set based on their estimated propensity score. Matching units on estimated propensity scores is not perfect and does not guarantee that matched treatment and control groups have good balance in relevant covariates. In addition, estimated propensity score helps to balance treatment and control group units on observed covariates, yet it cannot help balance the matched treatment and control group units on unobserved covariates, which can bring an element of potential bias to standard error estimate for sample ATT.

Lastly, the non-parametric bootstrap approach proposed above is computationally efficient. For both simulation studies above, using a Windows desktop computer with Intel(R) Xeon(R) Platinum 8-score CPU 2.6 GHZ and 32 Gigabyte RAM and the R Matching package, the maximum computation time for each sample of treatment and control groups is at maximum around 5 minutes across the different data scenarios while we implement certain parallel computing mechanisms into our code. With rapid advances in computing power and availability of distributed computing frameworks, it is quite possible that the computation time can be further reduced for the proposed bootstrap approach.

\section{Appendices}

\noindent\textbf{Appendix A. Relative size of overlap region}

\begin{figure}[!htb]
  \centering
  \includegraphics[keepaspectratio=true, height = 9.5 cm]{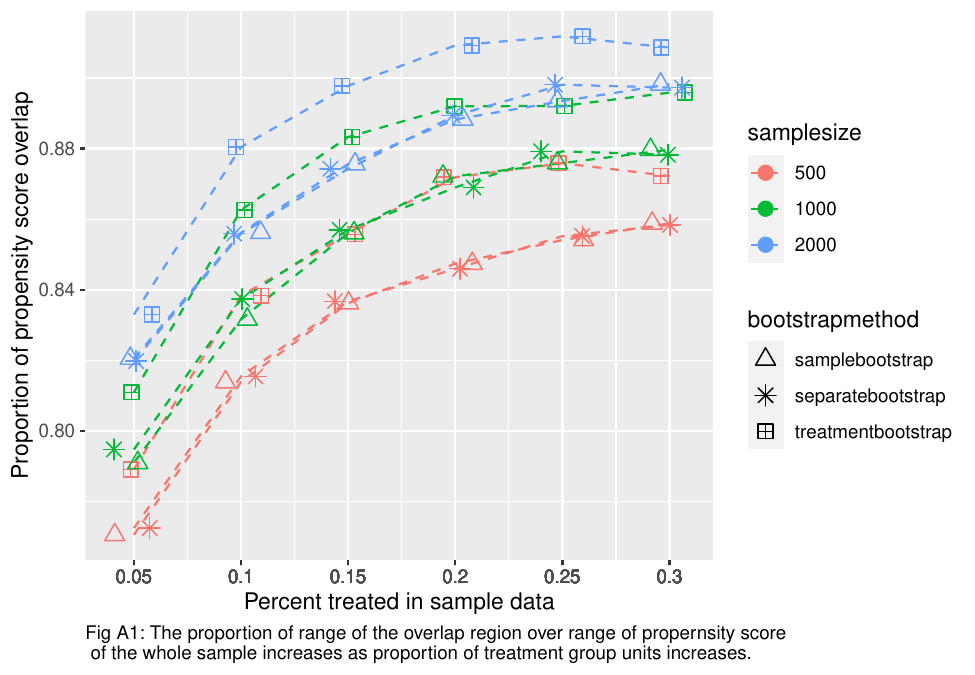}
\end{figure}

\noindent\textbf{Appendix B. Average number of potential matches for treatment units}

\begin{figure}[!htb]
  \centering
  \includegraphics[keepaspectratio=true, height = 9.5 cm]{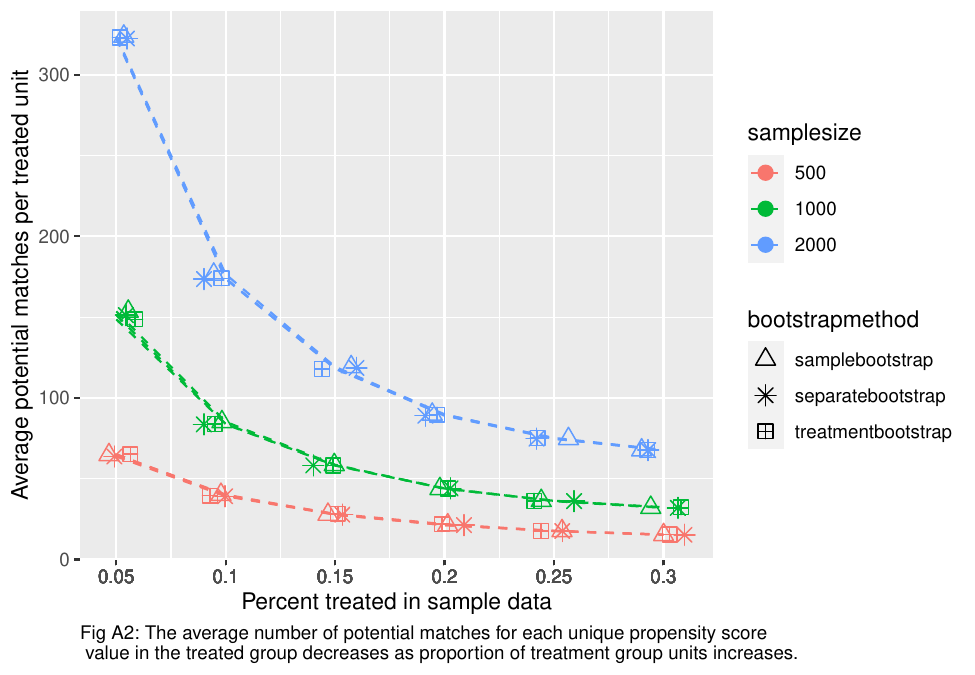}
\end{figure}


\begin{thebibliography}{99}

\bibitem[Abadie and Gardeazabal (2003)]{Abadie03}
Abadie, Alberto, and Javier, Gardeazabal. (2003). The Economic Costs of Conflict: A Case Study of the Basque Country. \emph{The American Economic Review}. 2003.

\bibitem[Abadie and Imbens (2002)]{Abadie02}
Abadie, Alberto and Imbens, Guido W. (2002).Simple and Bias-corrected Matching Estimators for average treatment effects. National Bureau of Economic Research 2002; NBER Working Paper Series(283).

\bibitem[Abadie and Imbens (2006)]{Abadie06}
Abadie, Alberto and Imbens, Guido W. (2006). Large Sample Properties of Matching Estimators for Average Treatment Effects.
\emph{Econometrica}, \emph{74}(1), 235–267.

\bibitem[Abadie and Imbens (2008)]{Abadie08}
Abadie, Alberto and Imbens, Guido W. (2008). Notes and comments on the failure of the bootstrap for matching estimators.
\emph{Econometrica}, \emph{76}(6), 1537–1557.

\bibitem[Abadie and Imbens (2011)]{Abadie11}
Abadie, Alberto and Guido W. Imbens. (2011). Bias-Corrected Matching Estimators for Average Treatment Effects, \emph{Journal of Business and Economic Statistics}, 29:1, 1-11

\bibitem[Abadie and Imbens (2012)]{Abadie12}
Alberto Abadie and Guido W. Imbens. (2012). A Martingale Representation for Matching Estimators, \emph{Journal of the American Statistical Association}, 107:498, 833-843.

\bibitem[Abadie and Imbens (2016)]{Abadie16}
Alberto, Abadie and Guido, W. Imbens. (2016). Matching on Estimated Propensity Score.\emph{Econometrica}, \emph{84}(2), 781-807.

\bibitem[Abadie, Diamond and Hainmueller (2010)]{Abadie10}
Abadie, Alberto, Diamond, Alexis and Hainmueller, Jens. (2010).
Synthetic Control Methods for Comparative Case Studies: Estimating the Effect
of California’s Tobacco Control Program.
\emph{Journal of the American Statistical Association}, \emph{105}(490), 493-505.

\bibitem[Abadie and Spiess (2022)]{Abadie22}
Abadie, Alberto and Spiess, Jann. (2022). Robust Post-Matching Inference, \emph{Journal of the American Statistical Association}, 117:538, 983-995.

\bibitem[Aldrich and Kage (2011)]{Aldrich11}
Aldrich, D., and Kage, R. (2011). Japanese Liberal Democratic Party Support and the Gender Gap: A New Approach. British Journal of Political Science, 41(4), 713-733.  Standard errors provided but no details.

\bibitem[Andrews, W.K. Donald (2000)]{Andrews2000}
Andrews, W.K. Donald. (2000).
Inconsistency of the bootstrap when a parameter is on the boundary of the parameter space.
\emph{Econometrica}, \emph{68}(2), 399-405.

\bibitem[Austin and Small (2014)]{Austin14}
Austin, C.Peter, and Small, S.Dylan. (2014).
The use of bootstrapping when using propensity-score matching without
replacement: a simulation study.
\emph{Statistics in Medicine}, \emph{33}, 4306-4319.

\bibitem[Austin and Stuart (2017)]{Austin17}
Austin, Peter and Stuart, EA. (2017). Estimating the effect of treatment on binary outcomes using full matching on the propensity score. \emph{Stat Methods Med Res}, \emph{26}(6), 2505-2525.

\bibitem[Austin (2022)]{Austin22}
Austin, PC. (2022). Bootstrap vs asymptotic variance estimation when using propensity score weighting with continuous and binary outcomes. \emph{Statistics in Medicine}. 41(22): 4426–4443. 

\bibitem[Boas and Hidalgo (2011)]{Boas11}
Boas, T.C. and Hidalgo, F.D. (2011). Controlling the Airwaves: Incumbency Advantage and Community Radio in Brazil. \emph{American Journal of Political Science} 55: 869-885. 

\bibitem[Bickel and Freedman (1981)]{Bickel81}
Bickel, J. Peter, and Freedman, David. (1981)
Some asymptotic theory for the bootstrap.
\emph{Annals of Statistics}, \emph{9}(6), 1196-1217.

\bibitem[Bodory et al. (2020)]{Bodory20}
Bodory,Hugo, Lorenzo Camponovo, Martin Huber and Michael Lechner (2020). The Finite Sample Performance of Inference Methods for Propensity Score Matching and Weighting Estimators. \emph{Journal of Business and Economic Statistics}, \emph{38}(1), 183-200.

\bibitem[Embens and Menzel. (2021)]{Embens21}
Guido Imbens. Konrad Menzel. "A causal bootstrap." Ann. Statist. 49 (3) 1460 - 1488, June 2021.

\bibitem[Cerulli (2014)]{Cerulli14}
Cerulli, Giovanni. (2014)
treatrew: A user-written command for estimating average treatment effects by reweighting on the propensity score. \emph{The Stata Journal}, \emph{14}(3), 541–561.

\bibitem[Christakis and Iwashyna (2003)]{Christakis03}
Christakis, Nicholas A. and Iwashyna, Theodore J.. (2003).
The Health Impact of Health Care on Families: A matched cohort study of
hospice use by decedents and mortality outcomes in surviving, widowed spouses.
\emph{Social Science and Medicine}, \emph{57}(3), 465–475.

\bibitem [Davison and Hinkley (1997)]{Davison97}
Davison, A.C. and Hinkley, D.V. (1997). \emph{Bootstrap Methods and their Application}.
Cambridge, UK ; New York : Cambridge University Press.

\bibitem[Dehejia (2002)]{Dehejia02}
Dehejia, Rajeev H. and Wahba, Sadek. (2002).
Propensity Score-matching methods for nonexperimental causal studies.
\emph{The Review of Economics and Statistics}, \emph{84}(1), 151–161.

\bibitem[Diamond and Sekhon (2013)]{Diamond13}
Diamond, A. and Sekhon, J.S. Geneticmatching for estimating causal effects: A generalmultivariate matching method for achieving balance in observational studies. (2013). \emph{Rev. Econ. Stat.} 95. 932–945.

\bibitem[Dynarski (2003)]{Dynarski03}
Dynarski, S. M. 2003. Does Aid Matter? Measuring the Effect of Student Aid on College Attendance and Completion. The \emph{American Economic Review.} 93, 279–288.

\bibitem[Efron (1979)]{Efron1979}
Efron, B. (1979).
Bootstrap Methods: Another Look at the Jackknife.
\emph{Ann. Stat}, \emph{7}(1), 1-26.

\bibitem [Fisher, R.A. (1935)]{Fisher35}
Fisher, R.A. (1935). \emph{The Design of Experiments}.
Oliver and Boyd. Edinburgh.

\bibitem[Galiani et al. (2005)]{Galiani05}
Galiani, Sebastian, Gertler, Paul and Schargrodsky, Ernesto. (2005).
Water for Life: The Impact of the Privatization of Water Services on Child Mortality.
\emph{Journal of Political Economy}, \emph{113}(1), 83–120.

\bibitem[Gerber and Green (2005)]{Green05}
Gerber, Alan S. and Green, Donald P. (2005).
Correction to Gerber and Green (2000), Replication of Disputed
Findings, and Reply to Imai (2005)
\emph{American Political Science Review}, \emph{99}(2), 301-313.

\bibitem[Hall and Marin (1988)]{Hall88}
Hall, Peter and Martin, Michael. (1988). On the bootstrap and two-sample problems.
\emph{Austral. J. Statist.} \emph{30A}, 179–192.

\bibitem[Hansen (2004)]{Hansen04}
Hansen, B.B. (2004). Full matching in an observational study of coaching for the SAT.
\emph{J. Amer. Stat. Assoc.}, \emph{99}, 609–618.

\bibitem[Henderson and Chatfield (2011)]{Henderson11}
Henderson, John and Sara, Chatfield. (2011).
\emph{The Journal of Politics} 73:3, 646-658.

\bibitem[Hill and Reiter (2006)]{Hill06}
Hill J, Reiter JP. (2006). Interval estimation for treatment effects using propensity score matching. \emph{Statistics in Medicine} 25(13):2230–2256.

\bibitem[Holland (1986)]{Holland86}
Holland, Paul W. (1986).
Statistics and Causal Inference.
\emph{Journal of the American Statistical Association}.

\bibitem[Hong and Park (2016)]{Hong16}
Hong, J., and Park, S. (2016). Factories for Votes? How Authoritarian Leaders Gain Popular Support Using Targeted Industrial Policy. British Journal of Political Science, 46(3), 501-527.

\bibitem[Imbens (2004)]{Imbens04}
Imbens GW. Nonparametric estimation of average treatment effects under exogeneity: a review. Review of Economics and Statistics 2004; 86:4–29. 

\bibitem[Imbens and Rubin (2015)]{Imbens2015}
Imbens, Guido and Rubin, Donald. (2015).
\emph{Causal inference for statistics, social, and biomedical sciences :
an introduction}. Cambridge University Press.

\bibitem[Imbens and Menzel (2021)]{Imbens2021}
Imbens, Guido and Menzel, Konrad. (2021). A Causal Bootstrap. \emph{The Annals of Statistics} 49(3), 1460-1488.

\bibitem[Kam and Palmer (2008)]{Kam08}
Cindy D. Kam and Carl L. Palmer. (2008). 
Reconsidering the Effects of Education on Political Participation.
\emph{The Journal of Politics} 70(3), 612-631.

\bibitem[Kocher et al (2011)]{Kocher11}
Kocher, M.A., Pepinsky, T.B. and Kalyvas, S.N. (2011). Aerial Bombing and Counterinsurgency in the Vietnam War. \emph{American Journal of Political Science} 55: 201-218. 

\bibitem[Lalonde (1986)]{Lalonde86}
Lalonde, Robert J. (1986).
Evaluating the econometric evaluations of training programs with experimental data.
\emph{Amer. Econ. Rev.}, \emph{76}(4), 604-620.

\bibitem[Mayer (2011)]{Mayer11}
Mayer, Alexander. 2011. Does Education Increase Political Participation?. \emph{The Journal of Politics} 73 (3): 633–645. 

\bibitem[Mozer et al (2020)]{Mozer20}
Mozer, R., Miratrix, L., Kaufman, A., and Jason Anastasopoulos, L. (2020). Matching with Text Data: An Experimental Evaluation of Methods for Matching Documents and of Measuring Match Quality. \emph{Political Analysis} 28(4), 445-468. 

\bibitem[Neyman (1923)]{Neyman23}
Neyman, J.N. On the application of probability theory to agricultural experiments. Essay on principles. Section 9 (1923), \emph{Stat. Sci.} 5 (1923 [1990]), pp. 463–480. reprint. Transl. by Dabrowska and Speed.

\bibitem[Otsu and Rai (2017)]{Otsu17}
Taisuke Otsu and Yoshiyasu Rai. (2017). Bootstrap Inference of Matching Estimators for Average Treatment Effects, \emph{Journal of the American Statistical Association}, 112 (520), 1720-1732. 

\bibitem[Rosenbaum and Rubin (1983)]{Rosenbaum83}
Rosenbaum, P.R. and Rubin, Donald B. (1983).
The central role of the propensity score in observational studies for causal effects.
\emph{Biometrika}, \emph{70}(1), 41-55.

\bibitem[Rosenbaum and Rubin (1985)]{Rosenbaum85}
Rosenbaum, P.R. and Donald B. Rubin (1985) Constructing a Control Group Using Multivariate Matched Sampling Methods That Incorporate the Propensity Score, The American Statistician, 39:1, 33-38.

\bibitem[Rosenbaum (2002)]{Rosenbaum02}
Rosenbaum, P.R. (2002). \emph{Observational Studies}. Springer. New York.

\bibitem[Rubin (1986)]{Rubin86}
Rubin, Donald B.  (1986).
Statistics and Causal Inference: Comment: Which Ifs Have Causal Answers.
\emph{Journal of the American Statistical Association}, \emph{81}(396), 961-962.

\bibitem[Rubin (1974)]{Rubin74}
Rubin, Donald B. (1974).
Estimating causal effects of treatments in randomized and nonrandomized studies.
\emph{J. Educ. Psychol}, \emph{66}(5), 688–701. 

\bibitem[Rubin (1978)]{Rubin78}
Rubin, Donald B.(1978).
Bayesian inference for causal effects: The role of randomization.
\emph{Ann. Stat}, \emph{6}(1), 34–58.

\bibitem[Rojas (2009)]{Rojas2009}
Rojas, Gabriel Montes. (2009). A note on the variance of average treatment effects estimators. 
\emph{Economics Bulletin}. \emph{29}(4), 2937-2943.

\bibitem[Singh (1981)]{Singh81}
Singh, Kesar. (1981). On the Asymptotic Accuracy of Efron's Bootstrap. \emph{Ann. Stat}, \emph{9}(6), 1187-1195.

\bibitem[Sekhon (2011)]{Sekhon11}
Sekhon, J. S. (2011). Multivariate and Propensity Score Matching Software with Automated Balance Optimization: The Matching package for R. Journal of Statistical Software, 42(7), 1–52.

\bibitem[Simmons and Hopkins (2005)]{Simmons05}
Simmons, B., and Hopkins, D. (2005). The Constraining Power of International Treaties: Theory and Methods. American Political Science Review, 99(4), 623-631.

\bibitem[Tu and Zhou (2002)]{Tu02}
Tu, Wanzhu, and Zhou, Xiao-Hua. (2002).
A Bootstrap Confidence Interval Procedure for the Treatment Effect Using Propensity
Score Subclassification.
\emph{Health Services and Outcomes Research Methodology}, {3}, 135:147.

\bibitem[Urban and Niebler (2014)]{Urban14}
Urban, C. and Niebler, S. (2014), Dollars on the Sidewalk: Should U.S. Presidential Candidates Advertise in Uncontested States?. \emph{American Journal of Political Science} 58: 322-336.

\end{thebibliography}
\end{document}